\documentclass[aps,nofootinbib,onecolumn,notitlepage,11pt]{revtex4-2}

\usepackage{amssymb, amsmath, amsthm, amsfonts}
\usepackage{graphicx}
\usepackage{color}
\usepackage{mathrsfs}
\usepackage{physics}
\usepackage{mathtools}
\usepackage{tikz}
\usepackage{faktor}
\usepackage{tikz-cd}
\usepackage{bm}
\usepackage[hidelinks]{hyperref}
\usepackage{csquotes}

\DeclareMathOperator{\E}{\mathbb{E}}
\newcommand{\lag}{\mathcal{L}}

\renewcommand{\grad}{\nabla}

\DeclareRobustCommand{\SkipTocEntry}[5]{}

\usepackage[UKenglish]{isodate}
\usepackage[UKenglish]{babel}

\begin{document}

\title{A WORKED EXAMPLE OF THE BAYESIAN MECHANICS OF CLASSICAL OBJECTS}

\author{Dalton A R Sakthivadivel \\ \href{https://darsakthi.github.io}{\url{darsakthi.github.io}}}
\affiliation{VERSES Research Lab, Los Angeles, CA, 90016, USA}
\affiliation{Department of Mathematics, Stony Brook University, Stony Brook, NY, USA}
\affiliation{Department of Physics and Astronomy, Stony Brook University, Stony Brook, NY, USA}

\date{\today}

\begin{abstract}

Bayesian mechanics is a new approach to studying the mathematics and physics of interacting stochastic processes. Here, we provide a worked example of a physical mechanics for classical objects, which derives from a simple application thereof. We summarise the current state of the art of Bayesian mechanics in doing so. We also give a sketch of its connections to classical chaos, owing to a particular $\mathcal{N}=2$ supersymmetry.

\end{abstract}

\maketitle



\tableofcontents

\section{Introduction}

Under the free energy principle \cite{afepfapp}, Bayesian mechanics is a new approach to studying the mathematics and physics of interacting stochastic processes. In essence, Bayesian mechanics is a particular sort of mathematical physics for coupled random dynamical systems, providing a mechanical theory for how the statistical properties of physical systems change in a space of Bayesian beliefs, based on how their physical properties change in space and time \cite{2012, parr, on-bmech}. Bayesian mechanics is particularly interested in systems with some notion of regularity, termed `self-evidencing' systems in previous literature.\footnote{This term of art originates in neuroscience \cite{howhy}, where the free energy principle has its origins, as an attempt to explain the physics and philosophy of learning in the human cortex\textemdash viewed as a Bayesian mechanical problem, a brain learning about an environment is one random dynamical system performing inference about another, with an aim towards the attractor states characteristic of allostasis \cite{control}. Note also that this paper uses `system' differently to \cite{afepfapp}, which says `particle' where we say system and `system' where we say agent-environment loop.} A self-evidencing system occupies an attractor in the state space, and has some set of stereotypical behaviours definitional of the sort of system it is. The key deliverable of Bayesian mechanics is that a random dynamical system is an estimator for statistics or parameters of another system to which it is coupled, making it an inference machine, and that we can understand attractors and phases in state space in virtue of these belief dynamics. 

Prosaically, the point of learning and representation in `genuine' inference agents, such as biological and artificial neural networks, is to mirror the environment within the agent's own structure\textemdash its tissue or weight configurations.\footnote{This remark has been paraphrased from Maxwell J D Ramstead.} In this sense, however, encoding data from the environment in an agential structure is simply a coupling between agent and environment. Indeed, the foregoing statement is simply a statement that a coupled random dynamical system is the preimage of some function into its environment, $\sigma$, instantiating some representation of its environment within its own structure. This is the `self' of the agent referred to as a coupled random dynamical system. Bayesian mechanics is thus a new set of techniques for understanding the relationship between statistical quantities and the physical dynamics encoding those quantities, or reflected by those quantities, in coupled random dynamical systems, with application to new areas of statistical mechanics like stochastic thermodynamics \cite{afepfapp, parr}. 

\newpage

Albeit conceptually powerful, the use of the free energy principle (FEP) and Bayesian mechanics to describe specific physical systems is rarely codified in the literature. There are examples where it reproduces known algorithms like various types of control \cite{beren, lance, eli}), as well as simple dissipative systems \cite{aguilera} and more complex systems exhibiting Lorentzian chaos \cite{conor}, but recent work has treated it as a purely formal position that systems constrain themselves to fall within acceptable regimes of certain existential variables, thus inducing such attractor structures in the state space. 

More specifically, in \cite{g-and-a, on-bmech}, Bayesian mechanics is introduced as the mechanics of beliefs\textemdash but it is challenging to determine the physical mechanics of systems carrying those beliefs. This would require solving difficult PDEs for non-equilibrium steady state densities in general, or else, equations of motion for internal states on the synchronous statistical manifold. Likewise, it is often claimed that the FEP is as simple and general as the principle of stationary action \cite{simpler}, and it can be sketched out how the Bayesian mechanics of internal states of classical objects might look \cite{afepfapp}. Nevertheless, there has yet to be a systematic investigation of even the Bayesian mechanics of classical physics, despite it being readily available due to recent formulations as a least action principle.

Here, inspired by remarks in \cite{on-bmech}, we give a worked example of the classical mechanics of objects with trivial (e.g., infinitely precise) belief dynamics\textemdash in a fairly direct sense, the simplest case of Bayesian mechanics \cite{g-and-a}. In doing so we provide a general formulation of classical physics for the working Bayesian mechanic. This paper not only surveys recent literature (including that of the author) and gives a constructive example of a Bayesian mechanical system, but ideally, will ground future discussions of the free energy principle even more solidly in conventional mathematics and physics. Of independent interest, it also serves as a derivation of classical physics from the principle of constrained maximum calibre, and recovers old results on supersymmetry theory in classical mechanics. 

\addtocontents{toc}{\SkipTocEntry}
\subsection*{Acknowledgements}

The author thanks Lancelot Da Costa, Karl J Friston, Brennan Klein, and Maxwell J D Ramstead for valuable conversations. This paper has benefitted from comments by Conor Heins and Igor V Ovchinnikov, as well members and collaborators of the \emph{Mathematically Structured Programming Group} at the University of Strathclyde, especially Matteo Capucci.



\section{Mechanics}

\subsection{Classical physics in one dimension}\label{class-mech}

The mechanics of classical objects are embodied by Newton's law that systems accelerate along force gradients, by precisely the direction and magnitude of the total force applied\textemdash no more, and no less. The derivation of this fact is simple. Let $K = \frac{1}{2}mv_t^2$, where $q_t$ is the position of some point mass at $t$ and its velocity is the time derivative $v_t = \dot q_t$, and $V$ be some scalar potential. Now define a path as a particular function\footnote{Imagine realisations of a random process as individual functions consisting of a drift added to a particular sequence of noise values; indeed, this is our motivation.} $q(t)$ (e.g., a parabolic path could be $q(t) = q_0 -\frac{1}{2}gt^2$, whilst a straight path could be $q(t) = q_0$), and introduce the temporary time variable $\tau$. Taking the action functional
\begin{equation}\label{classical-action}
S[q(t)] = \int_0^t \frac{1}{2} mv_\tau^2 - V(q_\tau) \dd{\tau}
\end{equation}
a path of \emph{least} action (or more generally, for which the action is stationary) obeys the equation 
\[
- \partial_{q} V = m \partial_{t} v
\]
by standard arguments in functional analysis. One can refer to \cite[Section 2]{on-bmech} and references therein for an overview; alternatively, see \cite{calder} for a more advanced pedagogical treatment, and \cite{gelfand, arnol2013mathematical} for mathematically sophisticated resources. This is Newton's second law, 
\begin{equation}\label{newton}
F = m a.
\end{equation}
Though appearing esoteric, this logical sequence simply expresses that a path of least action always follows \eqref{newton}\textemdash that is, a system always accelerates along a force gradient, never using extra energy by resisting or compounding that force. For an appropriate specification of forces $F$, we get various sorts of mechanics, like motion in gravitational fields or classical approximations of fluid flow (also called continuum mechanics). Given mass data, and initial conditions $\dot q(0)$ and $q(0)$ (along with other boundary conditions), we can produce dynamical trajectories for some particular system by actually using the law of motion given by Newtonian mechanics under the least action principle.

\subsection{The basics of Bayesian mechanics}\label{on-bmech-sec}

Bayesian mechanics can be seen as an account of the laws of motion deriving from the free energy principle, concerning how Bayesian beliefs\textemdash and hence, systems with beliefs, such as coupled random dynamical systems, which perform inference over the things to which they couple\textemdash behave under certain determinants of \emph{probabilistic} motion. Much like classical mechanics serves an account of systems that obey Newton's second law by minimising the classical action, Bayesian mechanics is an account of systems that engage in approximate Bayesian inference by minimising surprisal. In this and the following section, we will iterate over the main constructs of Bayesian mechanics at a progressively finer scale, unpacking the main insights from \cite{afepfapp, g-and-a, on-bmech}. The rest of the paper will contextualise those 250 or so pages within a worked example of physics in the classical world, furnishing new results along the way, such as the derivation of classical physics from the principle of constrained maximum entropy, as well as the (re-)introduction of supersymmetry as an explanation for classical chaos. The paper will proceed mostly linearly, first discussing the abstract principle that all coupled stochastic dynamical systems satisfying certain properties (e.g., the existence of a Markov blanket) follow, then discussing how that principle yields a law of motion for inferential systems, and then three examples of dynamics under that law of motion. (See \cite{on-bmech} for the original schematic of this distinction.)

The pure physics of the FEP arguably dates back to two landmark papers in the literature, \cite{afepfapp} and \cite{parr}. In \cite{afepfapp, parr}, and later in \cite{lance}, the idea was introduced that the FEP has gestured at a new sort of physics\textemdash one about the mechanics of Bayesian beliefs, and how they reflect the behaviour of systems carrying those beliefs. In \cite{g-and-a, on-bmech} it is discussed that one can understand this in the same sense as classical mechanics arises from the least action principle, or identically, that diffusion arises from the maximisation of entropy, with that least action principle being laid out in \cite{simpler} (to be formulated in detail in forthcoming work). Dually, we can understand our own beliefs about a system modelling its environment\textemdash or the system's model of itself\textemdash as being ruled by Bayesian mechanics, under the observation and updating rules which are a consequence thereof. A full deconstruction can be found in \cite{on-bmech}. 

What is central to Bayesian mechanics? Beginning with the most recent formulation in \cite{simpler}, the FEP is nothing but the least action principle applied to some surprisal\footnote{The empty argument $p(-)$ indicates our agnosticism about the input to $p$; it is important to note that the surprisal of different states constitutes a boundary condition yielding different sorts of dynamics \cite[Section 3]{on-bmech}.} $S = -\ln\{p(-)\}$, where the application of this `least surprisal principle' to specific objects determines the mechanical theory about those objects. Let a stochastic process $X$ under $p(x, t)$ with sample paths $\gamma$ be described by the It\=o stochastic differential equation
\begin{equation}\label{sde}
\dd{X_t} = f(X_t, t) \dd{t} + \sqrt{2D}\dd{W_t}.
\end{equation}
Here, $X_t$ is a random variable at $t$ and $f(X_t, t)$ is a vector field yielding the drift at any state $X_t$, which may itself change over time. Since $\dd{W_t}$ is a Wiener process, the expected state at some time $t$ is precisely $f(X_t, t)$. Immediately we arrive at an important qualification: by hypothesis, the ensuing discussion applies to cases where fluctuations are distributed as zero-mean Gaussian densities with constant variance $D$. This aligns with an ideal heat bath assumption.\footnote{The same limitation appears in both \cite{udo} and \cite{simpler}, which this work rests on. It is characteristic of most all of statistical physics, even frameworks valid away from equilibrium.}

Let $\omega_t = v_t - f(X_t, t)$, where $\omega_t$ can also be regarded as a label for the heuristic $\dot{W_t}$, be a fluctuation of any realisation of this flow at $t$. Note that a realisation $x(t) = \{X_s = x_s\}_{s\in 0:t}$ is (abusing types slightly) nothing but a sample path $\gamma$, and so, we have 
\begin{equation}\label{path-surp}
\dot \gamma_t - \E_{p(\gamma)}[\gamma]_t
\end{equation}
for $\omega_t$. A quadratic\footnote{Note that in the Stratonovich convention, more amenable to calculus on manifolds, there is an additional term in the Lagrangian $\lag$ indicated; see e.g. equation 15 here: \cite{path-int}. Physically, by ignoring that term, we have assumed perturbations to the flow have short characteristic timescales and are not `remembered' for long, and also, that the surprisal ought to be quadratic.} form $\lag(\omega_t)$ can naturally be defined on the tangent space to the state space, such that the surprisal is its integral along a given path $\gamma$ of $\lag$:
\[
S[\gamma] = \int_0^t \frac{1}{4D} \langle \omega_\tau, \omega_\tau \rangle \dd{\tau}.
\]
The Lagrangian, the integrand of $S[\gamma]$, is a function of noise. The surprisal of a path is then proportional to half its accumulated squared deviation from the expected flow $f(X_t,t)$, normalised by the scaling constant $\sqrt{2D}$. Morally, this is in the same sense as the classical action is proportional to half the square of the accumulated deviation of motion from a potential well \cite[Section 2]{on-bmech}. That this action equals\footnote{Note that the surprisal we go on to describe concerns the probability of an entire trajectory given a particular initial state\textemdash an entire path up to $t$.} the surprisal of a path $\gamma$ for a given initial condition, $p(x(t) \mid x_0)$, is a simple consequence of the path probability measure being defined as 
\begin{equation}\label{path-prob}
p(x(t) \mid x_0) = \exp{-\lambda S[\gamma]}
\end{equation}
in \cite{udo}, which is indeed the canonical definition of such an object in any abstract Wiener space \cite{oksendal, wiener}, and is the point of attack in \cite{simpler}. Such a `path-dependant surprisal' is referred to as the stochastic entropy by \cite{udo}, and is deeply related to statistics on the path space of a random walk (see \cite{machlup}, as well as \cite{bob} and related work on logarithmic heat kernels\footnote{The author thanks Robert W Neel for suggesting this point of discussion.}). This definition of the action is consistent with the quadratic form $\frac{1}{2} \langle v_t, v_t \rangle$ defined in classical mechanics. The action generates the Fokker-Planck equation for the probability of a state $x$ at $t$,
\[
\partial_t p(x, t) = - \partial_x [ f(x, t) p(x, t) ] + D \partial_{xx}p(x,t),
\]
by giving rise to a probability density over coordinate and time pairs. 

Define two random dynamical systems $\eta$ and $\mu$. In virtue of \eqref{path-surp}, the action of either system\textemdash and thus, ultimately, the surprisal\textemdash is parameterised by some modal or expected path. Suppose that $\eta$ and $\mu$ are coupled by some function $\sigma$, that one has an additional random dynamical system $b$ capturing the interactions between the two, and that conditional expectations $\hat \mu_{b,t} = \E_{p(\mu_t \mid b_t)}[ \mu_t ]$ and $\hat\eta_{b,t} = \E_{p(\eta_t\mid b_t)}[\eta_t ]$ exist; moreover, assume the statistics of the two processes can be distinguished, in the sense of being independent conditioned on $b_t$.\footnote{This framework degenerates in the case where $\sigma$ is the identity, but this case is vacuous, since it assumes $\eta$ is identical to $\mu$.} By construction, $\sigma(\hat\mu_{b,t}) = \hat\eta_{b,t}$. It is immediate that $\mu$ is an estimator for the statistics of $\eta$ \cite{g-and-a}. That is to say, in the case of random systems whose physical dynamics are coupled, these statistical quantities are also coupled, in a way that can be interpreted as the systems performing inference over each other. Recall that the path which minimises \eqref{path-surp} is the expected path. Our claim follows from the observation that the least surprising path of $\mu$ is the one which tracks the dynamics of $\eta$ across the synchronisation function $\sigma$, and \emph{vice-versa}; this means that systems that minimise their own surprisal must `know' something about their environment. Systems that minimise surprisal given a control parameter $\sigma^{-1}(\hat\eta_{b,t})$ are particularly good models of an environment, whilst systems that fluctuate with high probability are not. 

Hence, inference over an environment is equivalent in this sense to occupying unsurprising states. When we discuss models of `things' or systems, we are interested in states which are somehow `thing-like,' or systemic, i.e., attractors which define that thing, which that thing should not fluctuate too far away from \cite{g-and-a}. More broadly, general things which perform estimation must stay coherent in order to be an estimator (that is to say\textemdash we must have a $\mu$ to have a $\sigma$; see \cite{map-territory}); that this coherence is a necessary condition for synchronisation (read: estimation under a coupling), and that surprisal-minimisation follows from synchronisation, is simply the statement that things which exist reflect data about their environments. It is like saying that things which exist in the universe must obey the laws of physics as fundamental laws\textemdash when the wind blows and shakes a tree branch, on one reading, it is because the tree branch has inertia, but not enough to resist the force of the wind; on another, it is because the tree branch reduces its surprisal about the state of the world by modelling it.\footnote{The author owes this example to a similar remark stated to him by David I Spivak. For non-agential objects like the boughs of a tree, this latter view is artificially teleological\textemdash or perhaps even pure metaphor\textemdash and is referred to as `as if' inference. Though analogical in general, it has been argued that this does not constitute a failure of the model \cite[Section 4.3 and Remark 5.1]{g-and-a}; \cite[Section 6.3]{on-bmech}.} It is surprising not to follow these laws\textemdash for a tree branch to spontaneously resist the force of the wind entails a contradiction, which are usually surprising in the iron-clad realm of physical law. Either that branch no longer exists, in which case it can neither move with the wind nor be still, or, it has suddenly become much heavier than a tree branch. As a tautology, things that are surprising are things that ought not happen as we expect, like breaking the laws of physics. The aim of this paper is, in some sense, to see how insightful this analogy to forces and motion is.

We can understand this surprisal minimisation as the system holding an inferred model of the world parameterised by a mode and a variance, the `recognition density' $r(\eta; \hat\eta_b, \hat\vartheta_{\eta\mid b})$, arising from minimising a free energy functional. Suppose the system estimates the moments of the environment, such that a density $r(\eta \mid \mu)$ exists where $\sigma(\mu_b) = \eta_b$ such that $\sigma(\hat\mu_b) = \hat\eta_b$. Now consider the functional
\begin{equation}\label{vfe-1}
\int \ln\{r(\eta \mid \mu)\} r(\eta \mid \mu) \dd{\eta} - \int \ln\{p(\eta, b, \mu)\} r(\eta \mid \mu) \dd{\eta},
\end{equation}
the divergence between the variational model and the true joint density. This expands to 
\begin{equation}\label{vfe-2}
\int \ln\{r(\eta \mid \mu)\} r(\eta \mid \mu) \dd{\eta} - \int \ln\{p(\eta \mid b, \mu)\} r(\eta \mid \mu) \dd{\eta} - \ln\{p(b, \mu)\}.
\end{equation}
If $r(\eta \mid \mu) = p(\eta, b, \mu)$ then it also decomposes and the entire functional, including the surprisal of blanket and internal states, evaluates to zero; this is exact Bayesian inference. If the system estimates external states by modelling $r(\eta \mid \mu) = p(\eta \mid b, \mu) = p(\eta \mid b)$\textemdash which we can show \cite[Lemmas 4.1 and 4.2; Theorem 4.1]{g-and-a} occurs when $\mu = \hat\mu_b$ such that (again, abusing types slightly) $\sigma(\mu) = \hat\eta_b$\textemdash then the divergence in \eqref{vfe-2} vanishes. In other words, if the system estimates data about the environment by engaging in mode-matching, then an upper bound on $-\ln\{p(b, \mu)\}$ is minimised.

The principle of maximum calibre is a generalisation of maximum entropy to trajectories \cite{revmodphys}. Using this technology, we can construct the same model over trajectories of a process:
\begin{equation}\label{max-cal}
\int \ln\{r(\eta(t) \mid \mu(t))\} r(\eta(t) \mid \mu(t)) \dd{\eta(t)} - \int \ln\{p(\eta(t), b(t), \mu(t))\} r(\eta(t) \mid \mu(t)) \dd{\eta(t)},
\end{equation}
which appears exactly as expected. More about maximum calibre, and what has been called `path-tracking,' will be discussed later.

\subsection{A physics of beliefs}

The particular idea of inference under synchronisation in Section \ref{on-bmech-sec}\textemdash which packages together surprisal minimisation, estimation, and coupling into a modelling framework\textemdash is referred to as approximate Bayesian inference \cite[Theorem 4.1]{g-and-a}, and in particular, is referred to as `mode-matching' when these parameters are stationary: by minimising surprisal, the most likely state is $\sigma^{-1}(\hat\eta_{b,t})$. The \emph{a priori} assumption that systems minimise surprising events can be justified using large deviation principles \cite{touchette}, and as such, most any set of coupled random dynamical systems can be expected to perform approximate Bayesian inference of some sort. What is more striking is that two distinct (in the above sense) systems which estimate each other's statistics necessarily come equipped with a pair $(b_t, \sigma)$, chosen such that they fluctuate around each other's most likely states, and that any two systems with the right input/output flow (such that they fluctuate around each other's states) come with $(b_t, \sigma)$ such that they estimate each other's statistics \cite[Lemma 4.3]{g-and-a}.

Bayesian mechanics formulates changes in physical states as changes in probabilities estimated by $\eta$ and $\mu$ \cite{parr, on-bmech}. In this sense, Bayesian mechanics is continuous with the rest of statistical physics, and is merely a novel (and possibly more general) way of modelling how inference techniques like the principle of maximum entropy play a role in physics. Since we can understand the average state of $\mu$ as the preimage of $\sigma$, we can understand it as a parameter for the probabilities of states of $\eta$\textemdash in a sense we do this automatically, since we can understand a system $\mu$ existing a particular way in virtue of the likely states $\eta$ of the things interacting with it, causing or not causing particular $\mu$\textemdash and in so doing, we can relate $\mu$ to a belief mechanics by thinking of $\mu$ as encoding inferences about the assignment of probabilities to states of $\eta$. In simpler terms, systems exist in a particular way based on their environment. When we model a system, we automatically model it as modelling its environment by encoding this sort of statistical estimation in our model of the system. We expect a thing to exist as a particular `thing' based on whether or not it \emph{can} exist that way in a given environment. This places constraints on what $\sigma^{-1}(\hat\eta_{b,t})$ must be for a system to be `system-like' (e.g., stone-like, human-like, and so forth); dually, it informs what $\sigma(\hat\mu_{b,t})$ must be, given we have a particular system (e.g., humans require oxygen to breathe and cannot live for very long beneath water; stones do not mind.). In both cases we have a sort of allostatic attractor characteristic of the system. Besides the explicitly non-teleological\footnote{By \emph{teleological} we mean an explanation of the way something `is' which is based on the `purpose' of the system, the means by which a goal is achieved\textemdash here, the minimisation of surprisal\textemdash rather than its intrinsic nature\textemdash an enforced definition for itself, called an ontological potential \cite{g-and-a, on-bmech}. Indeed, one can explicitly contrast this with the normative role a set of constraints on system states plays. See \cite{mossio2017makes} or \cite{nahas} for recent reviews of the differences between these approaches in modelling self-organising systems, and how they also reinforce each other\textemdash namely, the minimisation of surprisal can be seen as allowing the system to meet some set of constraints as a goal, evincing an attractor in the state space. This appears to be one sort of statement of our duality \cite[Section 3, Section 6.2]{g-and-a}.} notion of the constraints or intended preimage which are definitional of a system, an important dual observation is that this parameterisation of likely internal and external states is one reading of Bayesian mechanics which is consistent with the idea of perception or estimation in Bayesian inference.

It is a deep result in \cite{afepfapp} that the existence of a system via the presence of a Markov blanket is the minimisation of surprisal, and in \cite{g-and-a} it is shown that we can understand that surprisal minimisation as the control of key existential or essential variables which are definitional of the system. Mathematically, as we stated above, it is equivalent to the fluctuations of a realisation of the system about that modal value parameterising the surprisal, and this is even more obvious in the form indicated in \eqref{path-prob}. Conceptually it is equivalent to asking that systems stay coherent if they are (i) at some optimal set-point despite their surroundings, or dually, (ii) encoding optimal beliefs about their surroundings. As we have already stated, there are two senses in which this is true: if a system estimates its environment, it is organised cohesively into the preimage of $\sigma$; dually, a system is an estimator only if it exists in a cohesive fashion to begin with. More complicated formulations of Bayesian mechanics update this statistical estimation to a sort of information gathering, where systems that are able to stay cohesive for longer and enforce an intended state learn enough about $\eta$ to change either $\hat\eta_{b,t}$ or the coupling $\sigma^{-1}(\hat\eta_{b,t})$, and hence maintain an intended $\hat\mu_{b,t}$. 

Viewing this as a maximum entropy problem (see \cite{on-bmech} or \cite{map-territory} for overviews, or \cite{g-and-a} for details), we have in \eqref{vfe-1} the constraint that the surprisal of the model of external states is, on average, zero\textemdash it is a perfect model \cite[Proposition 4.1]{g-and-a}. In \eqref{vfe-2}, we ask that the surprisal of the approximate model is on average no greater than the intrinsic surprisal of the system \cite[Proposition 4.2]{g-and-a}.

\subsection{A physics by beliefs}



We may also view such a constraint as applying to the internal states of the system; as stated, this is the control-theoretic view, which we are automatically sympathetic to in virtue of speaking of internal states as parameters of the free energy functional. 

We have leveraged the fact that, in virtue of being coupled, a value for $\mu$ parameterises a recognition density, and that this minimises a bound on surprisal, which arises from the system fitting a simpler parametric model $r(\eta; \sigma(\mu_b))$ to an approximate posterior density $p(\eta \mid b)$. Assume both densities are Gaussian or that $p(\eta\mid b)$ is well approximated by a Gaussian, a so-called Laplace approximation. Let $\Sigma$ synchronise variances, just as $\sigma$ did means or \emph{maximum a posteriori} estimates (what we have called modes). 
Consider the following: the KL divergence for the univariate Gaussians
\[
D_{\text{KL}}\left[r\!\left(\eta; \sigma(\mu_b), \Sigma(\vartheta_{\mu\mid b})\right) \| \, p(\eta; \hat\eta_b, \hat\vartheta_{\eta\mid b})\right]
\]
evaluates easily to
\[
\ln\!\left\{\frac{\hat\vartheta_{\eta\mid b}}{\Sigma(\vartheta_{\mu \mid b})}\right\} + \frac{\Sigma(\vartheta_{\mu\mid b})^2 + (\sigma(\mu_b) - \hat\eta_b)^2}{2\hat\vartheta_{\eta\mid b}^2} - \frac{1}{2},
\]
and this expression vanishes when $\mu = \hat\mu_b$ and $\vartheta_{\mu\mid b} = \hat\vartheta_{\mu\mid b}$, such that $\sigma$ and $\Sigma$ act on these parameters appropriately. Generalisations to the multivariate case follow easily. (Looking at this functional form, it is obvious that the variational free energy couples the two systems.) 

We might also observe that the system minimises a surprisal function of its own beliefs when the KL divergence vanishes, since $-\ln\{ r(\eta \mid \mu = \hat\mu_b) \} = (\sigma(\mu_b) - \hat\eta_b)^2 = 0$ in that case, where
\begin{equation}
\begin{aligned}
\E_r[-\ln\{ r(\eta \mid \mu) \}] &= \E_r\big[(\sigma(\mu_b) - \hat\eta_b)^2\big] \\ &\qquad= \Sigma(\vartheta_{\mu\mid b}) \\&\qquad= \E_{p(\eta\mid b)}\big[(\sigma(\mu_b) - \hat\eta_b)^2\big] = \E_{p(\eta\mid b)}[-\ln\{p(\eta \mid b)\}]
\end{aligned}
\label{block-eq}
\end{equation}
by construction (recovering Proposition 4.2 of \cite{g-and-a}). 

Moreover, the system minimises a surprisal function of its own states when the KL divergence vanishes, since $-\ln\{ p(\mu \mid \eta = \hat\eta_b) \} = (\mu_b - \sigma^{-1}(\hat\eta_b))^2 = 0$. Whilst subtler than it first appears,\footnote{See ``Adjoint Statistical Inferences,'' forthcoming, for an extended treatment of the functional form of this duality.} our na\"ive suspicion that the systems synchronise across both sides of the blanket is valid at the point of synchrony, since the roots of these two equations are the same. This allows us to view Bayesian mechanics dually, as a control problem, where the system maintains an unsurprising set-point $\hat\mu_b$. There is no guarantee of either being possible in general; the system may be constantly surprised if, for instance, it cannot occupy the expected value mirroring the environment (at which point it will cease to exist as it was, and transition to a new dynamical regime with whatever mean it \emph{can} occupy) or if the Gaussian approximation indicated here is a bad model of the environment (under which samples of $\eta$ are surprising). This `sleight of hand' is what lends the tautological truth that systems which exist do unsurprising things\textemdash and that coupled systems which exist synchronise their statistics\textemdash a more interesting, model-based interpretation.

As stated, if we have the two constraints
\begin{equation}\label{mean-constraint}
\E_{p(\mu\mid \eta)}[\mu \mid \eta] = \sigma^{-1}(\hat\eta_b) = \hat\mu_b
\end{equation}
and
\[
\E_{p(\mu \mid \eta)}\big[\mu - 
\E_{p(\mu \mid \eta)}[\mu \mid \eta]\big]^2 = \Sigma^{-1}(\hat\vartheta_{\eta\mid b}),
\]
where $\eta$ is now a choice of parameter, then we have a \emph{maximum entropy problem}: the expected surprisal of $\mu$ given $b$ is equal to the intended value of the expected deviation of $\mu$ given $b$ from the synchronised value of the parameter, which is the intended average. The foregoing statement that the average surprisal is non-zero, but the surprisal is zero for the average state, is that the distribution of $\mu$ given $b$ is a Gaussian, whose entropy (expected surprisal) is some intended variance (expected quadratic fluctuation): we have
\[
-\ln\{p(\mu \mid \eta)\} = (\mu - \sigma^{-1}(\hat\eta_b))^2
\]
for the stationary solution to that maximum entropy problem. Note that the dual of the constraint equation \eqref{mean-constraint}\textemdash i.e., that the mean of $r(\eta \mid \mu)$ is the same as the mean of $p(\eta \mid b)$,
\begin{equation}\label{synch-eq}
\E_{r(\eta\mid \mu)}[\eta \mid \mu] = \sigma(\hat\mu_b) = \hat\eta_b = \E_{p(\eta \mid b)}[\eta \mid b],
\end{equation}
is precisely the constraint that the mode of one density synchronises to the mode of the other. This reasoning also applies to the variance constraint following \eqref{mean-constraint}. 

In \eqref{block-eq} and \eqref{synch-eq}, we constrain the variance and mean of two different densities to be equal. Our observation that this is necessary and sufficient for the minimisation of variational free energy is nothing more than an instance of the information projection theorem. Note also a consequence of \eqref{block-eq}, that \emph{the average surprisal is generically non-zero}: the expectation of this surprisal, which is equivalent to a variance, is $\Sigma^{-1}(\hat\vartheta_{\eta\mid b})$. That is, if the variance is non-zero, then so too is the average surprisal. Instead, \emph{the surprisal of the average state is zero} and hence \emph{the ideal system, on average, minimises surprisal}, since the average state of that system is a surprisal minimiser. In parallel to this, the system explores the state space by fluctuating precisely how it must in order to sample the full suite of environmental states, $\hat\vartheta_{\eta\mid b}$. Since the path measure is formulated such that the average of fluctuations is the expected surprisal, and thus, the entropy, this is really a consequence that the entropy of a constrained maximum entropy distribution is whatever the average value of the constraint is, which is not generically zero (see \cite{dill-reply} for an argument along these lines). Prosaically, we could say that the average surprisal of the system is not minimised in general, but the surprisal is minimised by the system on average.

Finally, a remark on notation before we proceed further: notice that we have imposed the constraint that the internal state, \emph{sans} explicit consideration of the blanket state, is on average the synchronised average internal state \emph{given} a blanket state. Implicitly, we have incorporated a constraint that the input-output flows characterising the Markov blanket are aligned, such that blanket state $b$ is shared. In other words, this is not only a constraint that the system synchronises; we also fold in a constraint that the interface $b$ is shared, such that synchronisation is possible. The suggestion of this very point goes back to defining the synchronisation function as a function of two arguments, $\mu_b$, recapitulating the tensor-Hom adjunction in \cite[Lemma 4.3]{g-and-a}. Indeed, there it is shown that the synchronisation function exists if and only if the interface is shared, in which case, a partial function $\xi(b, -)$ of $\mu$ exists.\footnote{The same idea appears elsewhere in statistical physics in various disguises; see \cite{owen} or \cite{jaz} for detailed formulations of input-output composition.} 



This maximum entropy model is our belief about the state of the system whose beliefs are minimising free energy \cite{map-territory}. Since this is equivalent (in fact, dual) to a problem of variational free energy minimisation, the same laws of motion for belief updating that Bayesian mechanics defines apply to our own beliefs, and symmetrically, such laws are implemented by our beliefs about the systems being modelled. This makes Bayesian mechanics also a control problem: we model a system as leveraging its model\textemdash and these laws of motion in belief space\textemdash to engage in a kind of KL control, where the control parameter is the surprisal-minimising average state. Generalising, we could think of path-tracking \cite{on-bmech} as a path integral control problem given a reference trajectory $\hat\mu_b(t)$, conjectured in \cite{commentII} and discussed later in this paper (see also \cite{barp, eli} for recent work in this direction). 

Indeed, when we speak about control, we can dualise the above problem to ask that a system controls itself to maintain a system-like set-point. That is, we ask that the self-surprisal (the magnitude of fluctuations)
\[
-\ln\{p(\mu_{t} \mid \eta_t)\} =  (\mu_{t} - \sigma^{-1}(\hat\eta_{b,t}))^2
\]
is minimised, and moreover, that the accumulated surprisal 
\[
-\ln\{p(\mu(t) \mid \eta(t))\} =  \int_0^t (\mu_\tau - \sigma^{-1}(\hat\eta_{b,\tau}))^2
\]
is minimised. This inherits directly from our above discussion about path probability measures.

Bayesian mechanics leads to various types of approximate Bayesian inference, just like classical mechanics admits different applications of Newton's laws of motion (e.g., the continuum mechanics of fluid flow, or orbital mechanics for satellite motion). As stated, when that parameter is trivial, we have mode-matching; when it is dynamic, this is referred to as mode-tracking \cite{on-bmech}. When applied to beliefs about the trajectories of external and active states, we have a more general version of Bayesian mechanics only recently explored, including active inference \cite{barp}. This has been referred to as `path-tracking' in \cite{on-bmech}. The taxonomy described here exists in the same sense as minimising the action of the Lagrangian $\lag = K-V$ yields Newton's second law of motion, which can be applied to various sorts of systems when we know what sort of function $V$ is. We will work out in some detail what this taxonomy looks like in the world of classical physics.

In summary, Bayesian mechanics contains two key pieces of data: the surprisal action, and the synchrony map $\sigma$ (and thus, implicitly, a boundary). These data define beliefs and belief dynamics, respectively, and pair them with a characteristic geometry (information geometry \cite{amari}, the study of \emph{statistical manifolds}, or so-called `spaces of beliefs'). It is interesting that most physical theories are paired with a characteristic geometry in which they play out \cite{atiyah}, such as symplectic geometry in classical physics \cite{arnold}. Later in the paper we will point out the appearance of symplectic forms in Bayesian mechanics, which is notable given the analogy we draw.

\section{A general equation for Bayesian classical mechanics}\label{bayesian-class-mech}

We begin with a classical particle described by a position variable, $q$, at some time $t$. The mass of the particle will be denoted by $m$. The position plays the role of an internal state for the particle. The particle has an external environment interacting with it, whose states $\eta$ determine what forces are acting on it. This paper will build classical physics as the Bayesian mechanics of classical objects in an environment.

Why have we chosen classical physics as the setting of our exercise? Classical physics is well-established as the consequence of a particularly simple least action principle, with as many flavours of kinematics as there are forms of approximate Bayesian inference \cite[Section 3]{on-bmech} meaning that there is an opportunity to formulate straightforward least surprisal rules for this case. More interestingly, because the dynamics of classical systems are non-dissipative and exhibit periodic or chaotic motion, classical physics leads directly to challenging mathematical and physical situations which are of interest to describe. The characterisation of classical physics as a mechanics for infinitely precise beliefs makes it easier to handle some of this difficulty, which we will observe in Section \ref{g-theory-section}. Finally, if Bayesian mechanics is a physics whose boundary conditions are related to boundaries, we will see classical physics is particularly easy to formulate: the sparsity of the coupling is almost obvious.

Let $F_t$ be the vector of total force applied to the system at a time $t$, $\sum_i F_{i,t}$. The reception of an applied force is like a blanket state for the particle, which can couple to and affect internal states. The attentive reader will likely have noted that a force is not a state of the particle, but is an interaction of the environment with the states of the particle. Initially it may seem suspect that this Markov blanket is not part of the system, nor is it a physical state at all. It is perhaps a type error to identify a force with the measurement of that force, but beyond that, our construction is unproblematic for the following reasons: (i) the measurement of a force $F_t$ \emph{is} exactly a sensory state, (ii) a Markov blanket can be construed as simply that set of states which enforces the separation or distinguishability of two systems \cite{weak-MBs}, which a force certainly is, (iii) the failure of a force to map onto an internal state is precisely the failure for separation to be enforced, which is also when surprisal is high (see \cite{g-and-a}, where it is proven that the integrity of the boundary is equivalent to low system surprisal under a good variational model) meaning this functions as a Markov blanket regardless of its physicality. Moreover, simple objects have, in general, correspondingly simple Markov blankets.\footnote{See ``Path Integrals, Particular Kinds and Strange Things,'' forthcoming. See also \cite{james} for an argument that boundaries are information-theoretic objects; it follows that things with many states and a large amount of information to be transmitted need suitably complex boundaries, and things that do not, do not.} A single sensory state for a single classical point particle fits that bill.

That being stated, we consider $F_t$ itself to be like a sensory state of the particle. As such, we have an inverse synchrony map,
\[
\sigma^{-1} : \eta_t \to F_t \to q_t.
\]
The former map, $\bm\eta^{-1}$, merely sends external states of the world to the forces the world applies on objects, which will be done implicitly throughout the paper, as we provide worked examples with particular applied forces. Let $s$ be a temporary time variable. The latter map, $\bm\mu(F_{\eta,t}) = q_t$, is the solution to an integral equation determined by Newton's second law, 
\[
F(\eta, t) \mapsto \iint \frac{F(\eta, s)}{m} = q_t.
\]
In other words, the particular functional form for the coupling $\sigma$ we have used is one that sends the average acceleration of the system to the average force applied to the system,
\[
\sigma^{-1}(\eta_{F,t}) = q_{F,t}.
\]

The ideal path of internal states, which encodes an optimal (i.e., unsurprising) belief about what the system is being told to do by the environment, consists of precisely these $q_t$, for whom $\sigma(q_{F,t}) = \eta_{F,t}$. Note the consistency with more recent formulations of the FEP: for as long as there exists a particle, there exists some (possibly trivial) blanket distinguishing that particle from its environment along its path of evolution \cite{simpler, commentII}. We can argue that such a blanket exists\textemdash that any classical particle under the partition indicated above is sparsely coupled on the time-scale over which it exists\textemdash simply by pointing out that we can read off $F_t$ and need not consider $\eta_t$ to get internal states $q_t$. Physically, this is the intuitive statement that it is only a force that matters to motion, not what generated that force. Hence, by construction, for as long as a classical particle exists to be acted on, it has a blanket. See \cite{conor-comment, weak-MBs} for more on sparse coupling.

Just as we presume that the mechanics of beliefs should lead to physical mechanics (control), minimisation of the surprisal should lead to physical mechanics (Newton's laws of motion) and trajectories that abide by those mechanics. In establishing that physical mechanics follows from Bayesian mechanics, we first focus on beliefs \emph{about} internal states, and relate them to the beliefs \emph{carried by} internal states afterwards. The surprisal Lagrangian, $-\ln\{p(-)\}$, is applied either to $p(q_t)$ given $F_t$, or is applied symmetrically to the probability of $p(\eta_t)$ given $F_t$. We would like to see whether the minimisation of surprisal under $\sigma$ recovers classical physics in the context of Bayesian mechanics, so we try to minimise $-\ln\{p(q_t \mid \eta_{t})\}$ under the parameterisation $-\ln\{p(q_t ; \sigma^{-1}(\eta_{F,t}) )\}$. Here, we take a moment to remind ourselves that dualising the object in the Lagrangian to internal states gives us a physics of \emph{our} beliefs about the system, or the system's beliefs about itself, which is the duality indicated in \cite{g-and-a}. Contrast this with the surprisal Lagrangian on external states, which gives a physics of the system's beliefs about its environment, in \cite{simpler}. So, to set the stage, we change to the dual problem to \eqref{max-cal} (i.e., finding $p(\mu(t) \mid \eta(t))$ instead).

Under a noise injection, or else some uncertainty associated to the belief about a position at $t$, this becomes a problem of mapping means to means, in the sense of the approximate Bayesian inference lemma. We are primarily interested in the probability of deviating from the path of least action, such that our (state-wise) surprisal is a measure of 
\[
p(q_t; \sigma^{-1}(\hat \eta_{F,t})) = p(q_t; \hat q_{F,t}).
\]
Suppose the acceleration at a given time is constrained such that, on average, it obeys the forces being applied to it at that time. This is an instantaneous picture, licensing a non-dynamical application of Bayesian mechanics (note that we used that in Section \ref{class-mech} as well, where the instantaneous story behind derivatives licenses dropping the time variable from $q(t)$, a position in time). Denoting an expectation with $\E$ and neglecting subscripts, this gives us the equation
\begin{equation}\label{constraint}
\E_{p(q)}[q] = \iint \frac{F(\eta)}{m}.
\end{equation}
Suppose we also constrain the system to \emph{be} classical, in the sense of having infinite precision. This is a variance constraint, namely, that
\[
\E_{p(q)} \left[ \left( q - \E_{p(q)}[ q ] \right)^2 \right] = 0.
\]
When asking about paths, we ask that the accumulated variance of or along a path is also zero:
\begin{equation}\label{constraint-path}
\E_{p(q, t)} \left[ \int_0^t \left( q(\tau) - \E_{p(q, \tau)}[ q(\tau) ] \right)^2 \dd{\tau} \right] = 0.
\end{equation}
A similar law for the total squared displacement of a path has been used to produce Newton's laws from the principle of maximum path entropy (maximum calibre) before \cite{gutierrez}. Both of the above equations are simply a constraint that the optimal acceleration does not deviate from what the environment tells it to, which implies the approximate Bayesian inference lemma when under the additional constraint that the average state is the value of the synchronisation map \cite{g-and-a}. 

Here we have arrived at an important point: our belief about a system is an FEP-theoretic model of the system. We believe the system constrains itself to be the optimal parameter for some belief over what it is supposed to do in its environment. That parameter happens to be the least surprising internal state given some external state and a shared blanket state. This is what is meant in previous discussions about beliefs about internal states being dual to beliefs about external states. That is, in the sense that building a model of a system which constrains the system to exist at an ideal internal state is maximum entropy, maximising the entropy of our model under these constraints \emph{is} the `doing' of the free energy principle.\footnote{The idea that the maximum entropy principle under existential variables is the free energy principle under a synchronisation map was first suggested in \cite{andrews}. A proof of this can be found in \cite{g-and-a}.}

The optimal (i.e., least biased \cite{revmodphys}) belief under these two constraints can be derived from the principle of constrained maximum entropy, yielding
\begin{equation}\label{classical-path-prob-simple}
p(q) = \exp{ - \lambda_1 \abs{q - \lambda_2 \iint \frac{F(\eta)}{m}}^2 }
\end{equation}
with $\lambda_1, \lambda_2 > 0$ being Lagrange multipliers for the constraints indicated above. Considered dynamically, this equation can be given as a path probability density
\begin{equation}\label{gen-eq}
p(q, t) = \exp{ - \lambda_1(t) \int_0^t \abs{q(\tau) - \lambda_2(\tau) \iint \frac{F(\eta, s)}{m}}^2 \dd{\tau} }.
\end{equation}
When $\lambda_1^{-1}(t) \ll 1$ and $\lambda_2(t) = 1$ for all $t$, we can reproduce classical dynamics. In particular, in the limit $\lambda_1 \to \infty$, there is no uncertainty at all, and the most likely path under those constraints\textemdash the classical path of least action, by construction\textemdash is the \emph{only} path we lend any non-zero probability to. This is something like a classical limit for our path probability, in the same sense as taking $\hbar \to 0$ recovers classical mechanics from Feynman's path integral. Moreover, a reliable heuristic is that fluctuations in the theory are scaled by $D$ such that $\lambda_1 \propto D^{-1}$. The degree to which something can explore states within some allostatic bounds is precisely the variance under a Laplace approximation, yielding an important role for the Lagrange multipliers on the maximum entropy side of the story. This role is not necessarily evident in the FEP proper, nor in conceptual treatments of this duality which are divorced of process-level details. 

Finally, note that \eqref{gen-eq} is \eqref{path-prob} for a modal path given by \eqref{constraint}. This will be our general equation for Bayesian classical mechanics.\footnote{Note that, for a heuristic integral over an infinitesimal time, i.e., from $t$ to $\dd{t}$, \eqref{gen-eq} is equal to \eqref{classical-path-prob-simple}. This makes the equation with the time integral the general case. Intuitively, what we have noted is a statement that instantaneous variations in a path accumulate along the path as the path goes forward in time.}


\section{A question of quantum ontology}\label{quant-ont}

We began with the aim of showing that classical physics can be derived from Bayesian mechanics, by showing that deviations from a classical law of motion are surprising given a particular action functional, and that Bayesian mechanics describes the minimisation of surprisal. Do systems actually infer what their classical laws of motion are, and follow those inferences to avoid surprisal? More to the point: is there a less `just so' aspect of reproducing classical physics from the assumption that the least surprising trajectory of a system minimises the classical action? Can we do this without arbitrarily assuming the laws of classical physics and merely showing that unsurprising systems obey those laws? There is indeed a more concrete interpretation of this inference, one which makes the idea behind \eqref{gen-eq} more subtle.

In fact, what we have shown in the foregoing statements is that, under surprisal minimisation, a system takes a classical path when that path is the average. We can demonstrate that this has some further meaning by showing Bayesian mechanics is naturally derived from simpler arguments about the role of probability in quantum mechanics, such that the modal path of any fluctuating system is a classical path, and surprisal minimisation already exists in quantum mechanics. That is, we can derive Bayesian mechanics from the idea that a system is classical on average, just as we can derive classical mechanics from the idea that systems obey Bayesian mechanics for classical averages. This suggests a view that (i) systems with randomness do inference over their classical laws, and (ii) in the quantum setting we recover classical physics by doing inference. A more complete quantum physics manual for the Bayesian mechanic is to be written elsewhere. 

Begin from the supposition that classical equations of motion are asymptotics of quantum equations of motion, given by the empirical observation that we can measure classical effects more readily than quantum ones, but also, that classical equations of motion depend on parameters with underlying quantum effects, in such a way that quantum effects still bleed into the classical world when we `zoom in' to the extent that those parameters are no longer renormalised.\footnote{This is also referred to as an adiabatic approximation, and appears in semi-classical physics.} This leads directly to the \emph{correspondence principle}, a law of large numbers for quantum probabilities. The consequence of this classical `limit' is that the evolution of the most likely state of a quantum system gives us what we define as classical physics. Proven by Ehrenfest in 1927, we can rewrite this result (assuming $\abs{\partial_{tt} q(t)}$ is bounded above almost surely\textemdash physically, a thoroughly reasonable assumption) as 
\[
- \E_{p(q(t))}\big[\partial_q V(q)\big] = m \E_{p(q(t))} \big[ \partial_{tt} q(t) \big].
\]
So, assuming distance constraints like those above, such that we have a Gaussian measure or otherwise a Laplace approximation\textemdash a constraint which is quadratic in fluctuations\textemdash the most likely path ought to be a classical equation of motion. We will repeat this argument in Bayesian mechanical language, aided by the technology of the path integral.\footnote{We point the reader to \cite{hall} for details.}

In order to reproduce the idea that, probabilistically, quantum fluctuations are merely corrections to classical estimates, we take a Wiener measure where the variance of path probability\textemdash as it is given by the probability of the velocity on such a path\textemdash is scaled by some characteristic constant $\frac{m}{2\hbar}$,
\[
Z^{-1}\exp{-\frac{m}{2\hbar} \int_0^t \partial_s q(s)^2 \dd{s}}\dd{q(t)}.
\]
Note that, technically, we have Wick rotated our field theory. Note also that, appealing to the Trotter product formula, we can separate out the potential and assume it is zero everywhere on the support. 

As we remarked before, setting $\lambda_1 = \hbar$ and taking the scaling of quantum fluctuations to zero, and making use of the fact that fluctuations are precisely what contribute to the surprisal, we have a statement that in the quantum regime of Bayesian mechanical dynamics, the least surprising trajectory of the system is one that follows an overlying classical equation of motion. Indeed, under the WKB approximation, the most likely path in the path integral is the classical equation of motion of the field. Without quantum fluctuations about this classical solution we have classical physics, whereas in perturbative approximations to quantum mechanics, such as the quantum effective action, we add those quantum fluctuations in as higher order correction terms to a classical \emph{ansatz}.\footnote{This section assumes there is a \emph{unique} such classical solution to the system. Degenerate classical minima are handled by instanton theory, which we will not cover here. An excellent overview of the topic can be found in \cite{abcs}.}

So, we are now armed with two facts: basic physical observations suggest that classical paths are most likely paths, and, the canonical measure on paths is in terms of fluctuations about such a classical mode. We wish to see if the mathematical fact that this is the canonical description of path probability reflects the physical fact that classical equations of motion are the most likely paths of a system. Our formulation suggests that taking the limit $\hbar\to 0$ would be the right approach. Formally, this limit scales fluctuations to zero, revealing the most likely state as the one with constant velocity $v(0)$: a classical equation of motion under our identically zero potential, from where we derive no applied force and hence no acceleration.

Though we omit the proof, one can indeed prove that taking $\hbar\to0$ results in classical equations of motion. As noted by Feynman, this most likely path is what is most likely to be determined by observation (experiment), and thus determines the classical limit of the path integral. That this story\textemdash the most likely path is the classical one due to \emph{a priori} assumptions about the state and measurability of the world\textemdash implies the minimisation of surprisal, as well as following from it (as discussed in Section \ref{bayesian-class-mech}) is non-trivial.

What does it actually mean when we pass from `most likely' to `least surprising?' Least surprising to \emph{whom}? Certainly the experimenter\textemdash there is a sense in which the entire problem is dualised, and we are minimising the surprisal of \emph{our} beliefs about what a system does. That is, the two random dynamical systems coupled here are a quantum particle and an environment (including, perhaps, an external observer).

Do quantum particles do inference over where their classical paths are, organising themselves on average into the preimage of the average state of the observer, who expects to see a system follow a force applied? In the sense of estimating what that path is by taking a path which wastes the least energy (so to speak) in response to a force and in absence of quantum noise\textemdash and encoding such a path in their own dynamics, thereby inferring what another object `tells' them to do classically\textemdash they do. This makes Bayesian mechanics a useful way of modelling how a quantum system treats information and interactions with its environment,\footnote{We will refer the reader to \cite{james} and related references for more details about quantum information theory under the FEP.} defaulting to classical equations of motion on average. It is in this Bayesian mechanical sense that classical physics is a result of Bayesian inference in a quantum regime. It is not so trivial that, at the quantum level, such a particle is inferring what its classical trajectory\textemdash in modern parlance, inferring what its \emph{coherent state}\textemdash is, given some noise statistics and a macroscopic or higher-level force acting on it. It is this higher-level inference that organises the noisy, quantum system into a classical system.\footnote{The same phenomena has been demonstrated in self-organising soft matter systems \cite{neural}, suggesting this may be a promising line of research on collective excitations in condensed matter.}
 
More relevant to our case would be to return to the example of the boughs of a tree. A particle undergoing a force must minimise surprisal to stay cohesive, as we argued, and therefore can be read as inferring where the environment is directing it: what the environment is doing and how it is interacting with the particle. At the classical level, this is a concrete inference problem, wherein we try to find the \emph{maximum a posteriori} estimate of a density with thermal or quantum noise, furnishing the classical path `intended' by the environment. 

\section{The matching of modes}\label{ABIL}

This section begins a worked example of the typology described in \cite[Section 3]{on-bmech}. We begin with `mode-matching.' Mode-matching is the application of Bayesian mechanics to stationary objects which engage in approximate Bayesian inference \cite{afepfapp, g-and-a, on-bmech}. In this case, by definition of stationarity, the most likely internal state is fixed. Typically valid over only a brief time-scale\textemdash since nothing is stationary forever and nothing which is stationary and non-adaptive resists entropy for long\textemdash this is the simplest case of Bayesian mechanics. 

Inspired by \cite{g-and-a}, we formulate mode-matching under approximate Bayesian inference as internal states being constrained to be optimal parameters for a recognition density. Again, this is fully equivalent to the proper FEP. Using Theorem 4.2 (ibid), we can formulate the minimisation of surprisal applied to internal states $-\ln\{p(q \mid \eta_{F})\}$ as a demand that the log-probability equals some constraint on those internal states, with further precision-based minimisation possible over an ensemble of states. Here, that constraint is 
\[
S[q] = - \ln\{p(q \mid \eta)\} = \lambda_1\left( q - \iint \frac{F(\eta)}{m} \right)^2.
\]
Note the similarity to equation 3.5 in \cite{lance}. Under approximate Bayesian inference (and a further, but generically acceptable, Laplace assumption), the ideal state is the most likely state, which is the minimiser of this squared displacement.

The system we describe could be a stone performing inference over the cancelling of its gravitational pull and the normal force emanating from the ground, obeying  
\begin{equation}\label{sum-of-F}
\sum F = - F_g + F_N = 0.  
\end{equation}
In this case, the stone's acceleration is zero, and under the initial conditions $v(0) = 0$, $q(0)=0$, it goes straight to\textemdash and remains at\textemdash the mode $q = 0$. Referring back to the Introduction, the mode is a particular attractor which is a fixed point for the system. 
Indeed, for stationary initial conditions, the surprisal above is zero precisely when $(q - 0)^2 = 0$. 

Notably, this also means that for Bayesian mechanics to be consistent in the classical setting, it must (for stationary modes) imply Newton's first law\textemdash i.e., that for every applied force, there is a force of equal magnitude applied in the opposite direction. 

\section{The tracking of modes}\label{self-ev}


Mode-tracking can be summarised as the existence of a \emph{target} mode, i.e., a desired mode that systems are tracking towards, either for a finite time or constantly. This allows us to describe the most likely flow of autonomous states as a flow of beliefs \cite{afepfapp, parr, aguilera, commentI}, and involves the iteration of approximate Bayesian mechanics \cite{g-and-a}. Within mode-tracking we introduce two further distinctions, which is a more granular approach than the three-fold structure described in the Introduction: systems which track, but settle to, a mode, and systems which constantly chase a mode. The former is an example of a system that terminates in mode-matching, whilst the latter is an example of a system that is in constant motion. For both cases, we pass to an idea of dynamics, gesturing at an application of the principle of maximum calibre \cite{commentII}. 

\subsection{Terminal mode-matching}\label{terminal-mm}

Suppose the total force applied is dynamic, but eventually equilibrates. An example would be a stone tossed into the air, which travels through a gravitational field and eventually returns to the ground. The variant of \eqref{sum-of-F} corresponding to that motion is $F = -F_g$, and the solution of the integral equation \eqref{constraint} under that $F$ is 
\begin{equation}\label{LAP-path}
q(t) = q(0) + v(0)t - \frac{1}{2} g t^2
\end{equation}
where $g$ is the acceleration due to gravity. This equation has a steady state value where the mode $q(t) = 0$ and remains there, reached at some hitting time $t_\text{hit}$. For instance: for $v(0) = 0$ metres per second and $q(0) \approx 4.91$ metres above the ground, $t_\text{hit} \approx 1$ second. So, we can consider the full, dynamic-in-time force as 
\[
-F_g(t) + \mathbf{1}_\text{hit}(t) F_N(t), 
\]
where $\mathbf{1}(-)$ is an indicator function\textemdash the constant function equal to one on $t \geq t_\text{hit}$, and zero elsewhere. This terminates in the mode-matching explored in the previous section, since for all $t \geq t_{\text{hit}}$, we have $\sum F = 0$. Indeed, $q(t \geq t_{\text{hit}}) = 0$ by construction; this is a stationary mode.

By solving \eqref{constraint} as a constraint relation, we have actually asked that the average path obeys \eqref{LAP-path} and that no other path $q(t)$ deviates from that path. That is, we want $q(t) = \hat q(t)$ in the ideal case. As such, classical objects can be modelled as performing inference over the forces driving their motion, and given that they find known laws of motion unsurprising, get driven to the mode described in Section \ref{ABIL} by following \eqref{LAP-path}. For example\textemdash it would be surprising to the very fabric of space-time if a stone which had landed on the ground at some $t_{\text{hit}}$ were to spontaneously jump after. Thus, by obeying \eqref{LAP-path} and following classical laws of motion, the surprisal of motion is minimised. We will detail this below.

We begin by asking that $q(t)$ should equal $\hat q(t)$. Under our other constraint on the expected path\textemdash the solution to \eqref{constraint}, given above\textemdash eventually, $\hat q(t) = 0$ (in particular, for all $t \geq t_\text{hit}$). We could view this as dynamical inference more generally, or, the construction of a realisation of some path under a steady state density with mode zero, such that $\hat q(t) = 0$ after some convergence time (here, $t_{\text{hit}}$). In that case, the system goes directly to $\hat q$ as its kinetic energy `dissipates' into potential energy. 

In greater detail: a path consists of a list of states. Each such state is increasingly more likely as we approach the mode, and indeed, a mean-reverting process will go to a mode on average under a quadratic potential. As such, the average path taken by the system performs a gradient ascent on the probability density, or equivalently, a gradient descent on the surprisal. Note that we are not working in a path space here; rather, the path exists on a probability density, as a lift of a list of states, which is indeed simply a path\textemdash but, we don't speak about surprisal minimisation on such a path directly, instead speaking about the tendency of any path to go to a fixed point. Mathematically, this means that the motion of $q(t)$ is a gradient descent on $\abs{q(t) - \hat q(t)}^2$, with some convergence time $t_{\text{hit}},$ and $q(t) = \hat q(t) = 0$ for $t \geq t_{\text{hit}}.$ Since the logarithm of \eqref{classical-path-prob-simple} is this distance when $\lambda_2 = 1$, this is equivalently a minimisation of surprisal. That is, we have
\[
\partial_t q(t) = -\lambda_1 \grad \ln\{ p(q) \},
\]
such that the least surprising \emph{state} is the mode, and the system takes a path towards that mode. Moreover, the least surprising path to the least surprising state ought to traverse the distance from some initial $q(0)$ to $\hat q$ the quickest, which (for a fixed velocity) is the path given by a direct gradient descent on the distance. It would be of interest in future work to give a unified view of these approaches, i.e., to prove that under certain asymptotic conditions, least surprising paths are paths towards least surprising states.

Since $\lambda_1$ scales fluctuations, it is proportional to the inverse diffusion coefficient, or the covariance, more generally. Additionally, since there is no random motion and no opportunity to explore, the motion is described by a pure gradient descent, and so this yields one component of the Helmholtz decomposition discussed in \cite{afepfapp}. Note that the instantaneous Lagrange multiplier $\lambda$ plays a different role than the dynamical Lagrange multiplier $\lambda(t)$. In particular, the former selects states whilst the latter selects paths, a distinction that becomes important when we deal with paths towards least surprising states, as we have here. We still desire $\lambda_1(t)^{-1} \to 0$ as we did in Section \ref{quant-ont}, to reproduce classical path selection. In the Gaussian case this is precisely our uncertainty over paths, as we mentioned.

\subsection{Infinite mode-tracking}\label{infinite-mm}

This section will discuss itinerant objects whose gravitational field is such that the mode is never stationary\textemdash which we could call mode-chasing, as a sub-type of mode-tracking. Consider a planet in a gravitational potential equal in all directions: there is no stationary state, and hence, no meaningful mode, to the dynamics of this system. Na\"ively, there is no parameter through which to minimise surprisal. This does not mean the FEP cannot describe satellite motion. On the contrary\textemdash the application of the FEP to complicated systems is where it truly shines \cite{commentI}. Here, we can iterate self-evidencing to find out what constant motion dictated by a consistently dynamical principle might look like.

Like any classical system, in the absence of a force acting on it, a satellite system will continue to move on its trajectory. This is Newton's first law\textemdash the usual aphorism being, `a body in motion tends to stay in motion; a body at rest tends to stay at rest\textellipsis' with the phrase `\textellipsis{} unless acted upon by an outside force' often appended. As such, the hitting time formulation of Section \ref{terminal-mm} is no longer directly useful, but having \emph{no} hitting time certainly is. Note that the lack of a mode\textemdash but presence of circular, solenoidal flows\textemdash for true classical systems\footnote{That is, energetically conservative systems, or systems in the absence of dissipative forces. Although not a dissipative system, one can contrast this with the loss of energy \emph{of motion} that occurs upon colliding with the ground in Section \ref{terminal-mm}. It is consistent that such systems have a mode whilst unperturbed satellite motion does not.} is consistent with \cite{lance, g-and-a}.

Such a system has no dissipative component, since it is purely classical. This means the gradient descent describing the mode-matching dynamics we discussed in Section \ref{terminal-mm} degenerates, in the opposite sense as any exploratory component of the flow degenerated in that section. Here, we have a system which travels along a level set of a sphere of radius $r$, and in particular, travels such that the surprisal of states (parameterised by a stationary mode) is a non-zero constant along a path. By simple arguments in symplectic geometry\textemdash the geometric study of flows in classical physics \cite{arnold, ana}\textemdash flows which are level sets of some Lagrangian\footnote{Note that we typically consider a Hamiltonian, which is metric isomorphic to a Lagrangian.} and which admit radial coordinates are described by a skew-symmetric matrix. Indeed, level sets of the sphere centred on $\hat q$, projected down to the state space, are integral curves of the following equation:
\begin{equation}\label{symp-eq}
\partial_t \begin{bmatrix}
q_1(t) \\ q_2(t)
\end{bmatrix} = 
\begin{bmatrix}
0 &\nu\\
-\nu &0
\end{bmatrix}
\begin{bmatrix}
q_1(t) \\ q_2(t)
\end{bmatrix} + \frac{1}{\nu}\begin{bmatrix}
\hat q_1(t) \\ \hat q_2(t)
\end{bmatrix},
\end{equation}
where $\nu$ controls the system's frequency, or speed of travel along one such level set. Note that this system of equations corresponds to the second-order ordinary differential equation
\[
\partial_{tt} q(t) = -\nu^2 q(t)
\]
for either coordinate $q_1$ or $q_2$ (simply perform the matrix multiplication in \eqref{symp-eq}, take the derivative $\partial_{tt} q_i(t)$, and insert the expression for $\partial_t q_j(t)$ into the result) whose solution is
\begin{gather*}
q_1(t) = r\cos(\nu t) + r\sin(\nu t) + \hat q_1, \\
q_2(t) = r \sin(\nu t) - r \cos(\nu t) + \hat q_2
\end{gather*}
for some constant $r > 0$. Note also that 
\begin{equation}\label{surp-2}
-\grad\ln\{p(q, t)\} = 2\lambda_1(q(t) - \lambda_2 \hat q).
\end{equation}
As such, for an appropriate choice of $\lambda_1$ (namely, one half, or half the coefficients of the gradient descent operator, should it exist) and $\lambda_2$ (namely, $-1/\nu$), inserting \eqref{surp-2} component-wise into \eqref{symp-eq} yields 
\[
\partial_t q_i(t) = -\grad^i_j\ln\{p(q, t)\} = Q^i_j q_j(t) + \frac{1}{\nu} \hat q_i
\]
with $Q^1_2 = \nu$ and $Q^2_1 = -\nu$, which is exactly the other piece of the Helmholtz decomposition.

As we stated, the matrix operator indicated is skew-symmetric, and, the gradient on the sphere is locally orthogonal to these level sets\textemdash that is, moving on a level set does not change the gradient. Since there is no mode for these dynamics, we cannot describe a gradient descent on surprisal in the sense of Section \ref{terminal-mm}, but we can say that it is a gradient descent for which the value of the gradient is preserved, with velocity scaled by the matrix indicated. Future work will make the arguments given here\textemdash with respect to the Helmholtz decomposition being an artefact of the geometric nature of certain flows\textemdash more formal.

\section{Path-tracking, and a simple case of $G$-theory}\label{g-theory-section}

This section will progress the construction to more complex forms of path-tracking that are not amenable to the mode-based descriptions we discussed previously.

What has been called $G$-theory is the duality between maximum calibre and the genre of Bayesian mechanics applying to surprisal on paths \cite{on-bmech}, which we have begun to explore here. This pitches $G$-theory as the generalisation of the duality explored in \cite{g-and-a}, extending that construction to paths. In its full generality it is thought to accommodate descriptions of complex systems (like non-Markovian behaviours, moving attractors, and non-stationary statistics) more naturally and with greater fidelity than in the past. Some inspiration for this comes from the previously mentioned principle of maximum calibre, which does famously well on difficult problems in non-equilibrium statistical physics \cite{revmodphys, corey, annrevchem}. These results suggest that, whatever $G$-theory will prove to be, it will be a canonical modelling framework for complex systems.

Here we provide a simple example of this framework by formulating path-tracking, the least surprisal principle on paths\textemdash our third sort of approximate Bayesian inference\textemdash as an explicit problem of dynamical constraints. We then formulate a connection to chaos by examining the path-based nature of $G$-theory in greater detail. 



\subsection{Path-tracking}

Recall our results on mode-tracking in Section \ref{terminal-mm}. The construction there is obviously inelegant\textemdash besides formulating the path over the state space instead of doing proper dynamical inference, it exchanged the proper accumulated squared displacement in \eqref{gen-eq} with a less general instantaneous squared displacement at $t_{\text{hit}}$. Foreshadowing a more general extension to paths, the most natural formulation of this problem is readily seen as a gradient ascent on the path probability density. Under maximum calibre, our constraints lead to a probability density 
\begin{equation*}
\exp{ - \lambda_1(t) \abs{ q(t) - \hat q(t) }^2 },
\end{equation*}
where the expected path \eqref{LAP-path} is denoted by $\hat q(t)$, and $\abs{q(t) - \hat q(t)}^2$ is limitingly zero. For dynamic $F$, this is a moving Gaussian, with mode centred along the path for a given state-wise marginal. That is, it is centred on the list of $q$'s visited classically for a list of times, like a crest in the path space that runs directly over the intended states (and hence, a sort of mountain of probability for realisations flowing along the state space, concentrating them in that region). Path tracking is obvious in this situation\textemdash it appears to follow a gradient descent on the action, finding the most likely path by finding the summit at each $t$ of $p(q(t))$.

We can indeed still discuss a gradient descent in this case, but it is a functional gradient on the action $S$, such that the gradient descent is on deviations along a path, minimising fluctuations from the path of least action\textemdash and this is precisely the principle of least action, or of least surprisal, when the path of least action is the expected path (guaranteed by \eqref{path-surp}, as discussed in Section \ref{on-bmech-sec}).

Let $\delta q(t)$ be some first order variation of a path away from a path of stationary action at $t$ (see \cite[Figure 1]{on-bmech} for a depiction), which is in fact a realisation at $t$ of some fluctuation away from the expected path. Analytically, this means that we have 
\[
\delta q(t) =  - \hat\grad S[q(t)]
\]
where the kernel of the gradient in the path space, $\hat \grad$, is a path such that the system only changes to second order under variation\textemdash that is, it is a path $q(t)$ such that the distance between $q(t)$ and $\iint F(\eta, t)$ is minimised. This equation simply expresses that the path of least variation is the stationary, or expected, path. This means the system will most likely settle into an evolutionary regime that follows the expected path, which is least surprising\textemdash however, note this is simply a model of that process, since there are no such fluctuations in classical physics. Instead, we are interested purely in the zero point of the gradient.

Recall what the surprisal is in this case\textemdash the logarithm of \eqref{gen-eq} is merely the classical action. As such, the statement that systems evolve on stationary points of the classical action functional follows directly from a gradient descent on path surprisal given that systems follow forces. As such, the above equation reduces to
\begin{align*}
\delta q(t) = -\hat \grad \ln\{&p(q, t)\} \\ &= \hat \grad \left[\lambda_1(t) \int_0^t \abs{q(\tau) - \lambda_2(\tau) \iint \frac{F(\eta, s)}{m}}^2 \dd{\tau} \right],
\end{align*}
which yields 
\[
\delta q(t) = 0 \iff \partial_{tt} q(t) - \frac{F(\eta, t)}{m} = 0.
\]
Since our path space gradient on surprisal reproduces the Euler-Lagrange equation as the functional gradient of $S[q(t)]$, this is precisely classical mechanics. (As before, future work\textemdash here, regarding the Euler-Lagrange operator in Bayesian mechanics\textemdash should be done to make this perfectly rigorous.)

In the infinite mode-tracking case, we have something similar. 
For simplicity, we take the path of a satellite moving about a fixed central body of radius $r$ as a circle
\[
q_1(t)^2 + q_2(t)^2 = r^2.
\]
This is a constraint that the expected path is a circle of radius $r$, and that realisations of $q$ ought to have norm $r^2$. We will parameterise this as an expected path which is $\hat q(t) = [\hat q_1(t), \hat q_2(t)] = [r\cos(t), r\sin(t)]$. The surprisal Lagrangian measures precisely these deviations from a circle, 
\[
\big\langle q(t) - [r\cos(t), r\sin(t)], q(t) - [r\cos(t), r\sin(t)] \big\rangle.
\]
In this form it is even more apparent that our Lagrangian, the quadratic form defined in \eqref{path-surp}, is a metric on noise, $\omega = q - \hat q$.\footnote{In fact, we could choose to denote the Lagrangian as $g(\omega, \omega)$, for reasons of geometric significance.} Once more, it is exactly the distance of $q(t)$ from a circular path parameterised by $[r\cos(t), r\sin(t)]$. As for surprisal\textemdash again, given that systems move on geodesics through space-time, it would be surprising to see a system change its path to deviate from the curvature of space-time, and thus, to not follow the induced potential field. The final relation we derive is 
\[
\delta q(t) = 0 \iff q(t) - [r \cos(t), r\sin(t)] = 0.
\]
Given centripetal forces and the absence of tangential forces on a radial parameterisation of the circle, we can derive that 
\[
\partial_{tt} q(t) - C\frac{M}{r^2} = 0,
\]
where $C$ is determined to be Newton's gravitation constant; this equation then yields the acceleration for a system orbiting a central body of mass $M$ in circular fashion.

\subsection{A first idea of $G$-theory}

In Sections \ref{bayesian-class-mech} and \ref{quant-ont}, we introduced the idea that surprisal minimisation arose from solving for the Lagrange multiplier for a constraint that the most likely path was an on-shell trajectory, described by \eqref{constraint-path}. When this Lagrange multiplier is limitingly zero, and the particle does perfect inference over the forces being applied to it, we have classical mechanics. Here the uncertainty over both environment and system necessarily degenerated.

This curiosity\textemdash that Bayesian mechanics, when cast in the language of the principle of maximum calibre, naturally leads to a path integral representation of classical mechanics\textemdash is upgraded to a more interesting observation that, when viewed through the lens of a classical system interacting with an environment, a path probability density is the most informative about the system; in particular, that it is the most general way of understanding what an environment `tells' the system to do, and how that can be represented probabilistically as the system estimating those forces and following them so as to produce good (that is, unsurprising) inferences.

Although path-tracking is already a more elegant way of discussing the simple problems on display here, as expected, the problems the full generality of Bayesian mechanics seeks to provide solutions to are radically different than the simple Newtonian laws of motion investigated thus far. Moreover, to produce key identities in classical physics like the Euler-Lagrange equation, we practically began from where we wanted to end up: with the assumption that classical systems follow forces.\footnote{However\textemdash in defence of the author, and with reference to Section \ref{quant-ont}, what we \emph{really} did was say that the least surprising path of a system with quantum or statistical fluctuations is the one that obeys a classical equation of motion\textemdash a less trivial result.} Here, we aim to eventually formulate chaotic or itinerant systems under Bayesian mechanics, as has been done for earlier forms of the free energy principle \cite{conor}. A sketch of one such result will be found in this second subsection.


Let $\Gamma$ be the space of paths and $C(t)$ be a source for the field (this is merely an external field driving $x(t)$, and is generically comparable to an electric current). At the path of minimal surprisal, and under the demand that there is no other path possible, we have a Dirac measure over the classical path. Hence, the solution can trivially be transformed into the following path integral representation:
\begin{equation}\label{susy-path-integral}
Z[C] = \int_\Gamma \prod_t \delta(x_t - x_{t,{\text{cl}}}) e^{-\lambda_1(t) \int_0^t C(\tau) x(\tau) \dd{\tau}} \mathcal{D}x(t)
\end{equation}
where $x(t)_{\text{cl}}$ is the classical solution to the equations of motion of interest and the product of Dirac measures over intervals of $t$ of size $\varepsilon > 0$ is given (as such, note our paths are discretised). The term $\prod_t \delta(x_t - x_{t,{\text{cl}}})$ in the path integral enforces a weight of one for classical paths and a weight of zero for all others. Note that we assume anti-periodic boundary conditions in \eqref{susy-path-integral}.


Two standard tricks are to rewrite a Dirac measure in terms of a Jacobian determinant, and then to rewrite a determinant in terms of a path integral over `ghost fields.' These are anti-commuting variables that behave like auxiliary fermionic fields. Following the procedure described in \cite{gozzi}\textemdash which entails rewriting the on-shell trajectory $x(t)_{\text{cl}}$ in terms analogous to second-order variations $\dot\omega$, rewriting the Dirac measure on that function as a particular determinant, and then, introducing a pair of ghosts $\theta$ and $\bar\theta$\textemdash we have the following transformation of the Bayesian mechanical path integral:
\[
\int_{\tilde\Gamma} \exp{ i \int_0^t \xi(v_\tau - v_{\tau,\text{cl}}) +i \bar\theta \dot\omega\theta \dd{\tau}} \mathcal{D}x(t)\mathcal{D}\xi\mathcal{D}\theta\mathcal{D}\bar\theta
\]
where we have taken $C(t)$ to be identically zero (hence it has disappeared) and introduced a temporary variable $\xi$ after passing to imaginary variables. Note that 
\[
i\int_0^t \xi(v_\tau - v_{\tau,\text{cl}}) \dd{\tau} 
\]
reintroduces a term proportional to our surprisal Lagrangian (in Fourier variables), arising organically from defining what it means to be a classical path\textemdash and that there is an additional term
\[
-\int_0^t\bar\theta \dot\omega\theta \dd{\tau} 
\]
arising from the transformations on $\prod_t \delta(x_t - x_{t,{\text{cl}}})$ described. The latter term corresponds to a fermionic sector of our theory, as discussed; whilst the surprisal term defines a bosonic sector in contrast (this appears to be consistent with \cite{constraint-geometry}). Note also that, in spite of the appearance of an imaginary quantity in the path integral, there is no constant $\hbar.$ That ultimately preserves the classicality of this path integral.

Moreover, these ghost fields define a pair of supercharges; this is due to invariance under a pair of BRST transformations which relate bosonic and fermionic degrees of freedom, and generate a superalgebra under the commutator. This is the definition of a \emph{super}symmetry: a theory which does not change when we exchange fermionic particles for bosonic particles, and \emph{vice-versa}; hence, a theory which is symmetric under the action of one or more supercharges, which here are a pair of so-called BRST operators. Bayesian classical mechanics is thus an $\mathcal{N}=2$ supersymmetry theory.\footnote{We will suggest \cite{review} to the reader for a classic, if advanced, review.} This supersymmetry has a striking interpretation that gives us a glimpse of the power of $G$-theory: the ghost fields themselves appear to correspond to Jacobi fields, measuring the divergences in state space of classical trajectories with similar initial points\textemdash a typical metric for chaos which can easily be related to Lyapunov exponents (see \cite{graham1988lyapunov} for a worked example in the case of supersymmetric formulations of stochastic dynamics). The breaking of this supersymmetry, when two ground states can be produced, is a candidate geometric basis for non-ergodicity \cite[pages 388--389]{algebraic}.


For the Bayesian mechanic who does not usually carry supersymmetries in their toolbox, we can zoom out and say the following: the extra ghost fields\textemdash whose introduction is what yields a supersymmetric theory, and conversely, which arise from making the theory supersymmetric\textemdash naturally encode the sensitivity of motion to initial conditions. This means that Bayesian mechanics is potentially a canonical set of tools with which to model systems exhibiting classical chaos. Likewise, the breaking of this supersymmetry, where $\theta$ and $\bar\theta$ no longer deform one trajectory into another, corresponds to certain non-periodic, non-mixing motions; or, the non-ergodic regimes underlying much of complex and non-equilibrium dynamics. (Formally, this gives rise to an integrable system, which is necessarily non-ergodic.)

Far more work remains to be done on the nature of supersymmetric Bayesian mechanics, and especially its connections to chaos and the Bayesian gauge theory introduced in \cite{constraint-geometry, g-and-a, on-bmech}; however, for now we only note that this exciting connection to certain features of complexity is evidently a mere consequent of the use of $G$-theory. The theory described above is known to refine to more nuanced discussions of chaos and instantonic dynamics in supersymmetric stochastic processes \cite[see especially Section 5.4]{igor}, potentially allowing for new descriptions of complex systems.

\bibliographystyle{alpha}
\bibliography{main}

\end{document}